# Text data mining and data quality management for research information systems in the context of open data and open science

**Otmane Azeroual[123], Gunter Saake[2], Mohammad Abuosba[3], Joachim Schöpfel[4]**

[1]German Center for Higher Education Research and Science Studies (DZHW), Schützenstraße 6a, 10117 Berlin, Germany
[2]Otto-von-Guericke-University Magdeburg, Department of Computer Science, Institute for Technical and Business Information Systems - Database Research Group, P.O. Box 4120, 39106 Magdeburg, Germany
[3]University of Applied Sciences HTW Berlin, Department of Computer Science and Engineering, Wilhelminenhofstraße 75A, 12459 Berlin, Germany
[4]GERiiCO Laboratory, University of Lille 3, 59650 Villeneuve-d'Ascq, France

## Abstract

In the implementation and use of research information systems (RIS) in scientific institutions, text data mining and semantic technologies are a key technology for the meaningful use of large amounts of data. It is not the collection of data that is difficult, but the further processing and integration of the data in RIS. Data is usually not uniformly formatted and structured, such as texts and tables that can not be linked. These include various source systems with their different data formats such as project and publication databases, CERIF and RCD data model, etc.

Internal and external data sources continue to develop. On the one hand, they must be constantly synchronized and the results of the data links checked. On the other hand, the texts must be processed in natural language and certain information extracted. Using text data mining, the quality of the metadata is analyzed and this identifies the entities and general keywords. So that the user is supported in the search for interesting research information. The information age makes it easier to store huge amounts of data and increase the number of documents on the internet, in institutions' intranets, in newswires and blogs is overwhelming. Search engines should help to specifically open up these sources of information and make them usable for administrative and research purposes.

Against this backdrop, the aim of this paper is to provide an overview of text data mining techniques and the management of successful data quality for RIS in the context of open data and open science in scientific institutions and libraries, as well as to provide ideas for their application. In particular, solutions for the RIS will be presented.

# 1 Introduction

The flood of research information in higher education and research organizations is steadily growing. Research information systems (RIS) are used to support the collection, integration, processing, storage and presentation of research information. With the help of RIS it is possible to get an up-to-date overview of research activities, to process and manage information on scientific activities, publications, applications, projects as well as research data, press and media reports, etc. For the evaluation and interpretation of the research activities of a scientific organization, data quality and validation are crucial. In order to ensure the quality of RIS in the long term, it is recommended that data quality management (data profiling, data cleansing and data monitoring) be viewed as a holistic and continuous improvement process.

Data quality management becomes even more crucial in the new ecosystem of open data and open science, an “approach to research that is collaborative, transparent and accessible” (European Commission, 2017). The policy of open science has a direct impact of research management and research evaluation insofar henceforth they have to consider and assess new forms of research practice and aspects of openness. Moreover, the process of research evaluation is changing, shifting from a traditional top-down strategy towards more interaction, dialogue and co-production (Tatum, 2017).

The knowledge of research activities and research results is becoming an increasingly important factor for the success of an organization. Yet, this knowledge requires, alongside quantitative data, more and more qualitative information extracted from a large variety of heterogeneous and often unstructured sources on the web, such as data and text files in digital libraries, open repositories, social networks and other platforms. For this reason, the new technology of *text mining* or *text data mining* of the rapid growth of digital, unstructured and semi-structured internal and external data is becoming increasingly important as a scientific method.

Text mining is a technique for analyzing documents or texts and extracting new knowledge unknown to the user. Thus, this developed technology is relevant for all scientific institutions and libraries that produce texts (such as editorial offices and publishers), edit text archives (libraries) or in which large amounts of text are constantly being generated. The aim of text mining techniques is to uncover interesting information or patterns in unstructured textual documents by allowing them to process the vast amount of natural language words and textures, and to handle the treatment of uncertain and fuzzy data. Text mining is a new area of research to solve problems of information overload through the use of data mining techniques, machine learning, computational linguistics (natural language processing), information retrieval, and knowledge management.

The aim of this paper is to present the technology of text data mining and the data quality management and to show how both can improve the performance of RIS, especially in the context of open data and open science.

## 2 Research information systems

A current research information system (CRIS) or research information system (RIS) is a *database management* or a *specialized federated information system*, that provides a continually updated, comprehensive directory of researchers and research activities (like for example, employees, projects, publications, patents, cooperation partners, awards and conferences etc.) of an organization. It serves to make the portfolio of research activities more visible and thus provide an overview for all interested or users.

The selection of specifications to be included in the RIS is based on the German Research Core Dataset (RCD). The RCD is a standard for the collection, provision and exchange of research information. The RIS in the context of the RCD is the instrument for the uniform definition and structuring of research information. To reach reliable data quality and optimized reuse of research information in and out of the institutions and libraries of distributed and heterogeneous databases, it is important to establish a high degree of connectivity through RCD for the effective exchange of research information in order to avoid duplication and multiple input of research information and the resulting sources of error. In addition to improving data quality, the RCD can greatly simplify work and reporting processes.

The building blocks of a RIS architecture are distinguished in three-stage processes (*data collection* and *data integration*, *data storage* and *data presentation*) (as illustrated in Fig. 1).

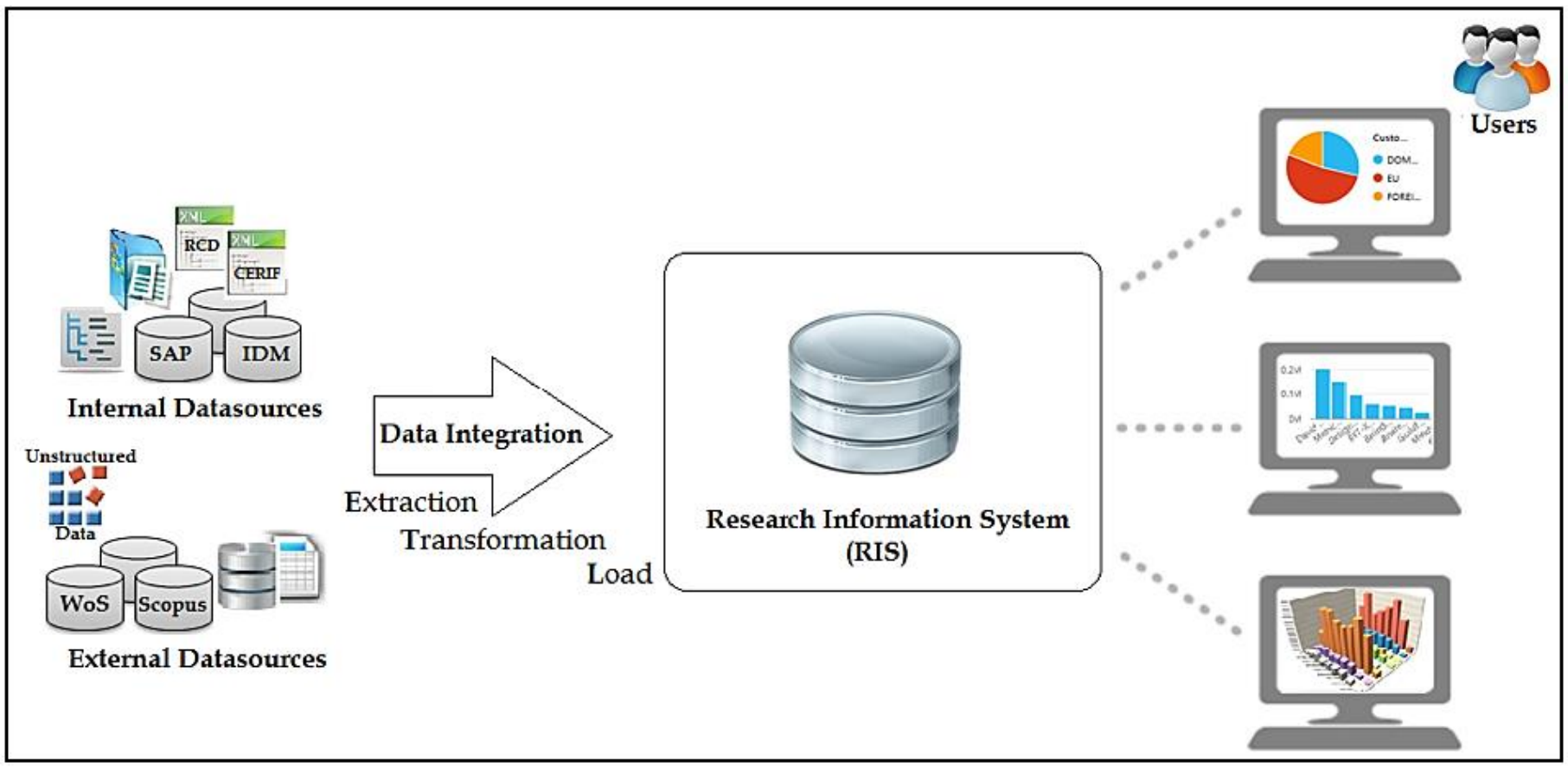

**Figure 1.** *Architecture of RIS (Azeroual et al. 2018c, d)*

The ***data collection*** contains the internal and external data sources. The RIS collects information about the research activities and research results associated with institutions and their scientists by automatically synchronizing existing data volumes with various data sources, so that this qualified research information is on hand to the management as a better basis for decision-making. For an automated data import from existing systems, a connection of internal as well as external application systems can be realized. These application systems, which ideally can be included to collect research information, are the identity management system and the campus management system as internal systems, as well as public publication databases and project databases. In the case of publications, these include, for example, Web of Science, Scopus or PubMed and the financial system (e.g. SAP) of third-party funds management for the data on third-party funded projects and the personnel management system (e.g. PAISY) for information about the scientific staff.

Offers for the standardized collection, provision and exchange of research information in RIS are the Research Core Dataset (RCD) data model and the Common European Research Information Format (CERIF) data model.

The ***data integration*** transfers, processes and compresses the required data from source systems (different information systems with different data formats and structures), which is described as the Extraction, Transformation and Load (ETL) process. In the extraction phase, data is extracted and pulled out from various source systems or source documents and provided for further processing steps in the input layer of the RIS. The transformation phase has the task to prepare the data for the loading process and to clean it up in order to convert the data extracts into a uniform internal format. The cleansing of incorrect source data is absolutely necessary and will be detected and corrected by plausibility checks. This data cleansing is done in four steps: filtering, harmonization, aggregation, and enrichment. In the next and last phase, the data is loaded into the RIS, there they are stored in a structured and normalized.

The aim of the ETL process is to ensure that the prepared data can be stored efficiently and persistently in data storage.

The ***data storage*** includes the RIS and its applications that merge, manage and analyze data held at the underlying level. If the data is present in the RIS, the reports required for evaluation can be set up.

In the ***data presentation*** is the target group specific preparation and presentation of the analysis results for the end user (decision-makers). These are made available as reports using the business intelligence tools. Beside to various reporting options, portals and websites of the organizations can also be filled here.

## 3 Standardization of research information

To ensure the aggregation, standardized collection and interchangeability of research information and to be able to integrate as many decentralized stocks of research information as possible, there are (inter)national standards for supporting RIS.

Two measures for the standardization of research information are the well-known international standard CERIF of the European organization euroCRIS[1] and Research Core Dataset (RCD)[2], which the German Council of Science and Humanities recommended in 2013. To support and advise the higher education institutions and non-university research institutions of the German science system on the implementation on the RCD, a corresponding helpdesk was established at the German Center for Higher Education Research and Science Studies (DZHW)[3] in Berlin, which answers questions about the content and technical of the RCD and provides an information base and cooperate events and workshops on the RCD.

- CERIF is a format recommended by the EU for its members that provides descriptions for research information (such as *persons*, *projects*, *organizations*, *publications*, *patents*, *products*, *funding*, *equipment* etc.). The CERIF model[4] can serve as a basis for RIS as an exchange format among various systems of research information.
- The implementation of the RCD at higher education institutions and non-university research institutions in Germany takes place on a voluntary basis. The RCD is a young data standard for research activities and considers it to be more effective and simpler for the complex and inconsistent reporting process for all research stakeholders from the German science system, as well as to reduce the collection of data from research activities and to facilitate the comparability of research data. The RCD defines core data that should be kept in aggregated form by all higher education institutions and non-university research institutions for various documentation and reporting processes. These include the focus areas of *(a) employees, (b) support for young researchers, (c) third-party funded projects, (d) patents and spin-offs, (e) publications and (f) research infrastructures* (see Biesenbender and Hornbostel, 2016 for details on the structure and logic of the RCD). In order to make the data compatible and interoperable at the international level, the technical RCD data model[5] offered was developed on the basis of the CERIF format.

The following Fig. 2 illustrates the example of the XML schema as a data exchange format for CERIF and RCD.

---

[1] https://www.eurocris.org/

[2] https://www.kerndatensatz-forschung.de/

[3] https://www.dzhw.eu

[4] See https://www.eurocris.org/cerif/main-features-cerif

[5] See https://kerndatensatz-forschung.de/version1/technisches_datenmodell/

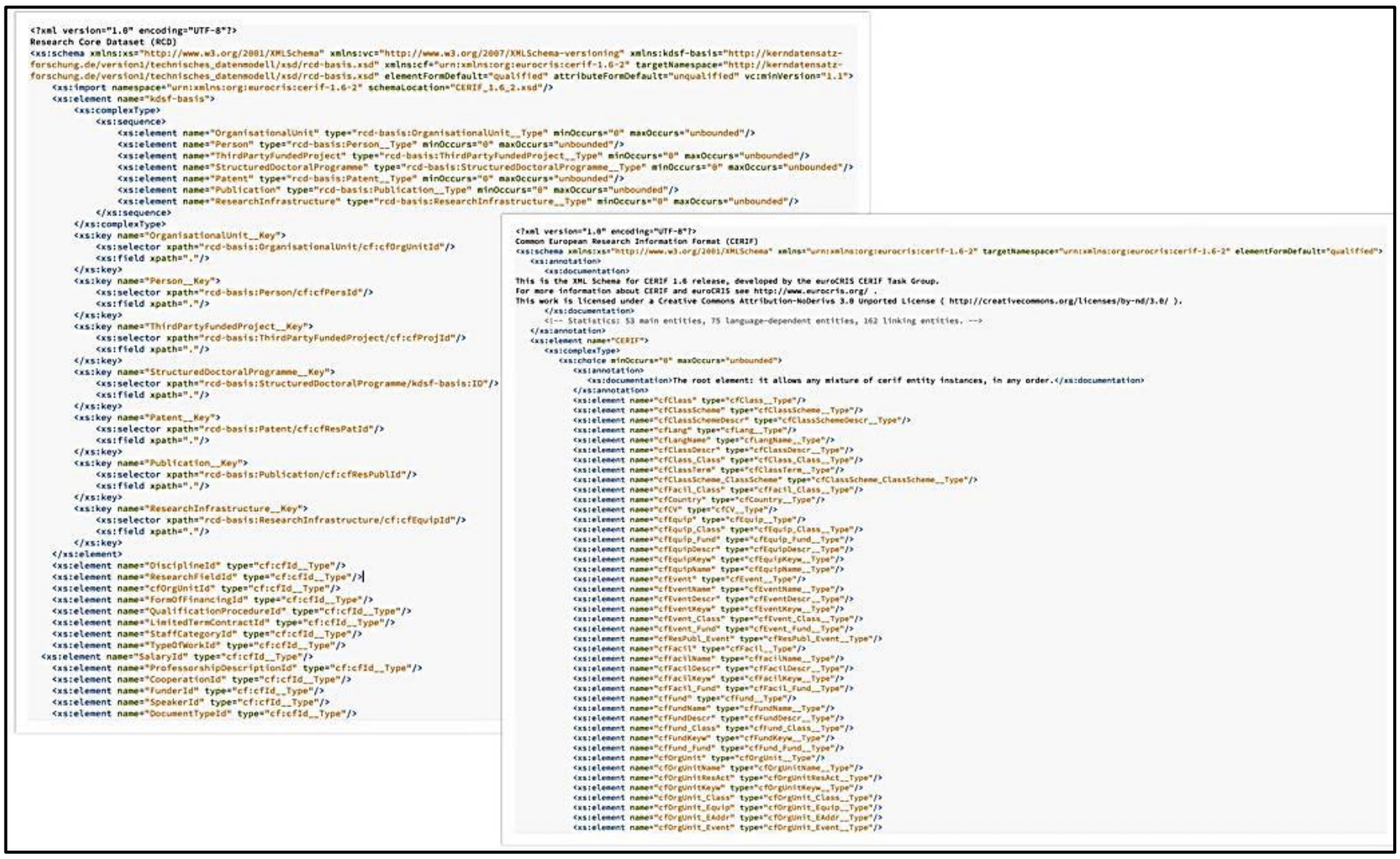

**Figure 2.** *XML schema for the CERIF and RCD standard*

Applying these two standards will benefit a common information infrastructure for both institutions and researchers. In order to use more benefits in the management of research information, for example by exchanging data, developments in standardization should be taken into account when building or selecting a RIS. The European portal OpenAIRE[6] is a good example. Its main goal is the wide and open dissemination of research results funded by the European Commission, based on the feed from deposits in institutional repositories and OpenAIRE's Zenodo "catch-all repository" hosted by CERN. OpenAIRE collects metadata from a variety of data sources, like open repositories, data archives and RIS. The interoperability of OpenAIRE and local RIS requires the implementation of standards like CERIF, RCD XML and OAI-PMH to guarantee high quality information interchange[7].

## 4 Research information systems and open science

Open science is an umbrella concept for a wide range of activities including open access publishing, open data, open peer review and citizen science; it goes hand in hand with research integrity and requires legal and ethical awareness on the part of researchers (European Commission, 2017). The French National Plan for Open Science defines open science as the "practice of making research publications and data freely available (...) drives scientific progress (and) fosters scientific integrity and people's trust in science" (MESRI, 2018). Open science policy impacts scientific practice in several ways and on different levels, and RIS must be able to monitor and evaluate these new features.

[6] Open Access Infrastructure for Research in Europe www.openaire.eu/

[7] See the *OpenAIRE Guidelines for CRIS Managers* on Zenodo https://zenodo.org/record/17065

A recent report written by the EC "Working Group on Rewards under Open Science" provides a matrix for the evaluation of research careers fully acknowledging open science practices (European Commission, 2017). This matrix OS-CAM provides a comprehensive framework that can be used to develop evaluation systems for various situations and contexts. It proposes a large number of possible evaluation criteria for the assessment of six domains and 24 open science activities, i.e. datasets and research results, risk management, peer review, knowledge exchange, teaching and even personal qualities.

OS-CAM tries to describe doing open science on the individual level, and it illustrates the multiple dimensions and complexity of what is called "open science". On the institutional level, the Harvard Open Access Project (HOAP)[8] from the Berkman Klein Center may provide another framework for the assessment of open science policies and services in universities, research organisations and other scientific structures.

RIS are not designed for a given science policy but are (or at least should be) "policy agnostic", able to comply with traditional research strategies as well as with the emerging new ecosystem of open science. Of course, insofar this new ecosystem changes the way research is done and modifies the way it is evaluated, it will have an immediate and direct impact on the RIS themselves. For instance, the French National Plan calls for the recognition of open science in assessments of researchers and institutions and requests to prioritise quality over quantity when evaluating research (MESRI, 2018).

Regarding the impact on RIS, we can distinguish three aspects.

1. Information sources: The RIS must be able to process data from new information sources, beyond the traditional data providers and outside academia, in order to monitor and evaluate new configurations of openness, involving other people, institutions and activities like NGOs, social networks, corporate companies etc. The reliability of these sources may be uncertain, such as the quality of the data; probably this means to adopt a strategy similar to the "big data" approach in the corporate sector.
2. Results: Open science changes the way how research is valued, and open science monitoring requires other metrics than the traditional H-index, impact factor article counting etc. "Evaluation outcomes aimed at present and future role of openness in research" (Tatum, 2017). Thus RIS must be able to produce open science related knowledge in a reliable way, representing societal impact (such as altmetrics[9]), openness (such as the SPARC open access spectrum[10]) or other qualitative information about participation, networking etc.

---

[8] https://cyber.harvard.edu/research/hoap

[9] https://altmetrics.org/manifesto/

[10] https://sparcopen.org/our-work/howopenisit/

3. Process: Finally, RIS must take into account the transformation on the research evaluation process itself. As already mentioned above, research evaluation is shifting from a traditional top-down strategy towards more interaction, dialogue and co-production (Tatum, 2017), and this transformation takes place on all levels, on the individual (career assessment, grant awarding), local (institutional), organisational and national level. For the RIS, this means more transparency and more participation from the very beginning; both are critical factors for the success of a RIS, especially in the environment of open science (Schöpfel, 2015). Also, its implementation should be an opportunity to make researchers aware of the new challenges (de Castro, 2018). In other words, a closed RIS in an environment of open science makes no sense.

Open data means data sharing and data reuse, and this applies also to research information. How do RIS produce reusable data? How do they reuse data produced from other systems? And how do they make research information available as open as possible? And do they, really? These questions are all related to standards and policy.

Returning to quality, (Azeroual et al. 2018c) highlight the need for high quality data produced by RIS is an essential consideration for evaluating research activities of all institutions. Because of the increasing heterogeneity of data input, of new evaluation processes and of the requirement of standard, machine-readable output (interoperability with other systems and infrastructures, cf. FAIR guiding principles), the quality especially of metadata becomes a crucial asset of the development and administration of RIS (Simons, 2017).

The impact of open science on the implementation of a RIS has been described by different case studies, with special attention to the role of academic libraries an information professionals (see for instance Clements and Proven, 2015 and Brennan, 2018). In the following chapters, our paper will focus on the metadata problem, especially when related to unstructured and semi-structured data. How can RIS produce and process high quality output based on low quality input?

# 5 Text data mining

## 5.1 Definition

For over thirty years, text mining has been a relevant topic in research, although at the beginning it was only used in a few disciplines (Upshall, 2014). The term text mining is a special form of data mining and is referred to as "*text data mining*" or "*text analytics*". The text data mining refers to computer-based methods for the semantic analysis of texts, which support the automatic or semi-automatic structuring of texts, especially very large amounts of texts (Heyer et al. 2008). In particular, text data mining aims to identify and extract knowledge that is implicit in the text that the user of the information system does not know.

## 5.2 Metadata – handling unstructured and semi-structured data

Higher education institutions and non-university research institutions that manage their research information with larger data sets in RIS face the chal-lenge because their information is not available in structured data but in semi-structured or unstructured data. The unstructured data comprises, among others, text documents or text publications in Word or PDFs, memos, emails, RSS feeds, wikis, images, video or audio data etc. Semi-structured data include, for example, CERIF or RCD XML or HTML files. Such data can be a time-consuming burden for RIS staff in institutions and libraries to seek and thus make bad decisions, because the unstructured and semi-structured data are stored in different formats and designated by different technologies. To solve this problem, metadata, which is data about data, should be stored with unstructured or semi-structured data. By implementing text data mining techniques, for instance classification and clustering algorithms (see below), metadata about research activities can be generated and organized (Seth et al. 2009 and Bhanuse et al. 2016).

### 5.3 Integrating text data mining into research information systems

Today, information on scientific activities and outcomes of science organizations is collected, maintained and published in a variety of forms and vast amounts of data through RIS. However, these are mostly unstructured in various forms and media and present a great challenge. A traditional, manual approach is not enough to investigate large amounts of data or document collections from various meta-databases (e.g. PubMed, Scopus etc.). Therefore, text data mining techniques can be helpful to analyze large numbers of text publications, library-produced data or even the content of blogs or tweets. The following Fig. 3 shows a framework for supporting text data mining in RIS.

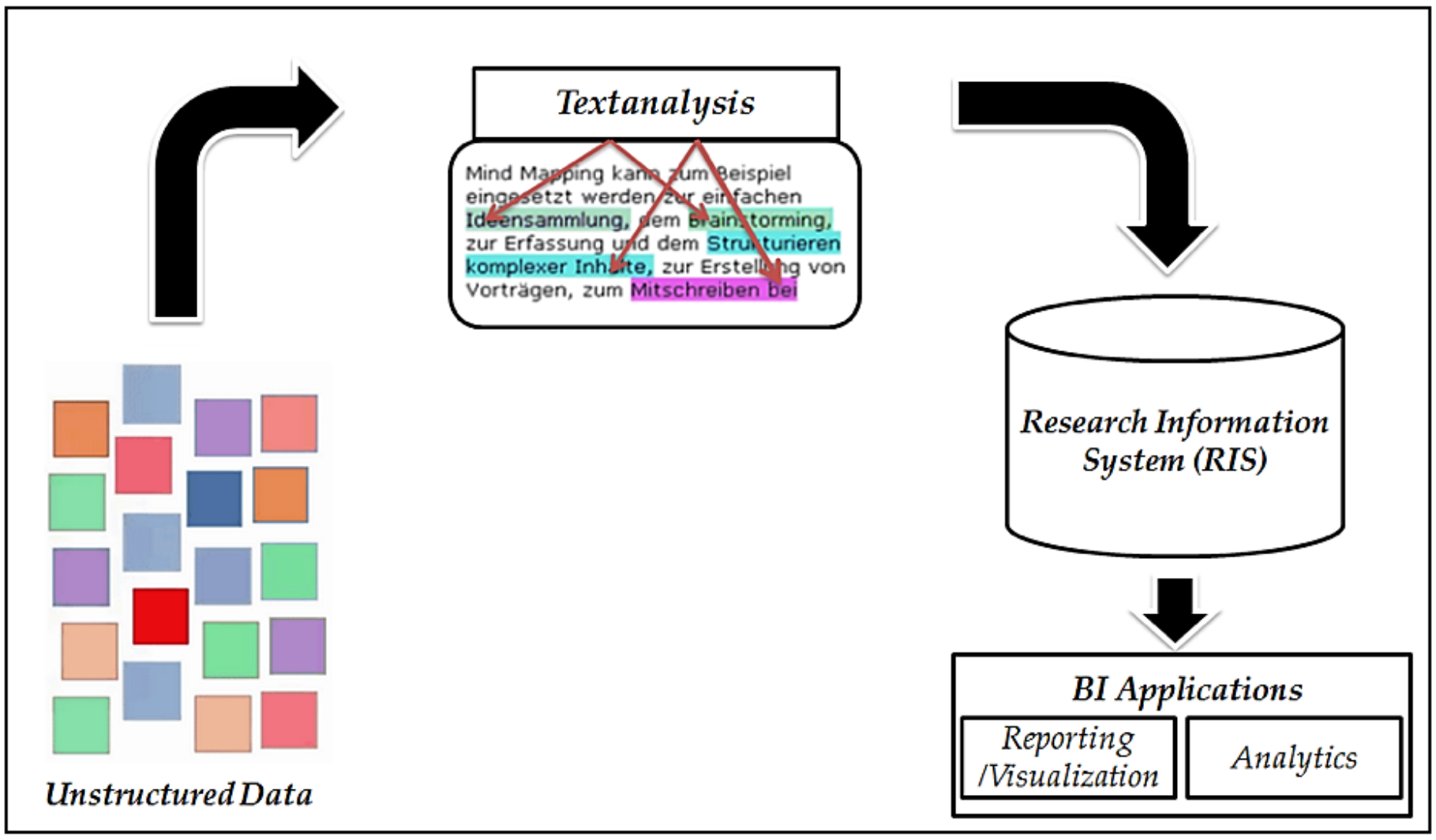

**Figure 3.** *Overview of support text data mining framework to RIS*

The peculiarities of text data mining are the structure of the data as well as the origin of the data. The data type to be examined is unstructured text from internal or external sources.

In the following section, the most important techniques of text data mining for supporting RIS in the context of open data and open science are presented in an overview for the institutions and libraries that use them.

### 5.4 Techniques in text data mining to support research information systems

Because of the vast amount of textual data available in RIS from various internal and external sources, it is not possible to read them completely and correctly and perform a manual analysis. In this case, text data mining is needed to convert text into data that can then be examined using various analysis techniques. These techniques are classification, clustering, NLP, information extraction and information retrieval.

***Document classification / Categorization***

Classification assigns research information in different categories. For example, in this way publication texts can be classified according to their content-related focus (e.g. computer science, economics, politics etc.). Classification procedures identify patterns to make statements about objects based on existing research information. First, the already existing objects are grouped according to their known feature or behavior regarding the problems to be analyzed in different classes. From this set of objects a classification model is developed, with which one can then predict the class affiliation of a new object. In the classification analysis methods of artificial intelligence or decision-tree-oriented methods can be used.

***Document clustering***

Clustering consolidates documents into clusters. Clusters of documents can be used to quickly find similar documents and detect duplicates. Unlike classification, clustering does not use a predefined set of terms or taxonomies that are used to group the documents (Liao et al. 2012). Instead, groups are created based on the document features that appear in a set of documents to be clustered.

***Natural Language Processing (NLP)***

In order to extract research information from natural language texts, it is necessary to use methods of natural language processing. NLP is an application of computational linguistics, which is responsible for the communication and interaction of humans and computers. The goal is for computers to understand human statements, from single words to complex texts, and to process their content or based on given circumstances, generating natural language as output (Chopra et al. 2013). For the automatic semantic analysis of texts, the use of ontologies is an essential tool. Ontologies are controlled, structured vocabularies that contain entities of a domain and their relationships to each other. They are necessary to make texts interpretable for the computer (Carstensen et al. 2010). Ontologies are used for named entity recognition (NER), which identifies proper names of entities (such as persons, organizations, etc.) in texts. This can be used to automatically keyword texts.

***Information extraction***

In information extraction (IE), research information is found in textual documents and translated into structured form (that is, identifying specific text passages and assigning them to structured objects). For example, research information about persons, organizations, publications, projects, etc. can be found and correlated.

*Information retrieval*

Information retrieval (IR) is the finding or retrieving of relevant documents that provide answers to specific questions. Here, texts are searched for using keyword requests, and a common example is keyword searches using search engines like Google, Yahoo, or Bing. Thus, the amount of relevant documents should be found by means of keywords and presented to the user in a result set.

Many institutions are still limited in their RIS still on the pure data collection and often go cumbersome ways to make it usable for themselves. With these steps of text data mining techniques, all institutions can better analyze and understand their stored texts in order to make appropriate decisions.

## 5.5 A model for research information processing

To assist the scientific institutions and libraries, the following developed process could be used to extract new and interesting information from different sources of unstructured documents using the techniques of text data mining and to derive new conclusions (see Fig. 4). In carrying out these featured text data mining techniques, there are a variety of supported commercial open source software solutions that offer multiple text data mining techniques. Nowadays, a developer, but no longer an expert, has to be in order to harness the potential of texts.

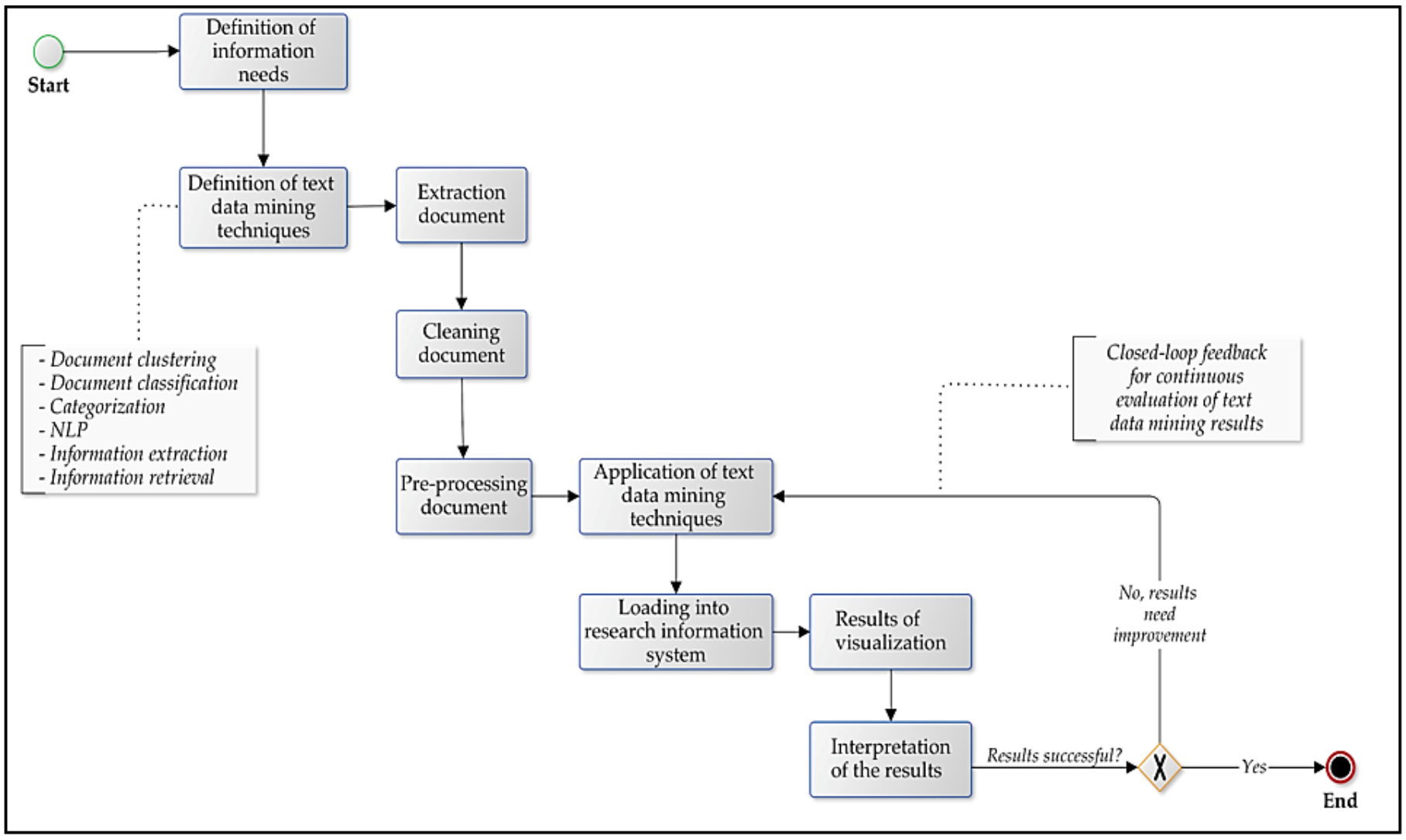

**Figure 4.** *Process of knowledge discovery using text data mining techniques*

Text data mining techniques can be used not only in scientific institutions, but also by libraries to improve their services, such as improving search and support in keyword coding.

# 6 Data quality management as a success factor for research information systems

In the scientific institutions and libraries, the management places the highest demands on the quality of the research information to be processed and evaluated. One of the biggest hurdles in securing data quality is typically the heterogeneity of the various internal and external data sources. Owning the right technology is just one way of ensuring data quality in RIS. Much more important to success is the use of the right methodology. This can be achieved by using the following data quality management as a cycle with its three phases, which must be run continuously to achieve a sustainable improvement and increase in data quality in RIS (as shown in Fig. 5).

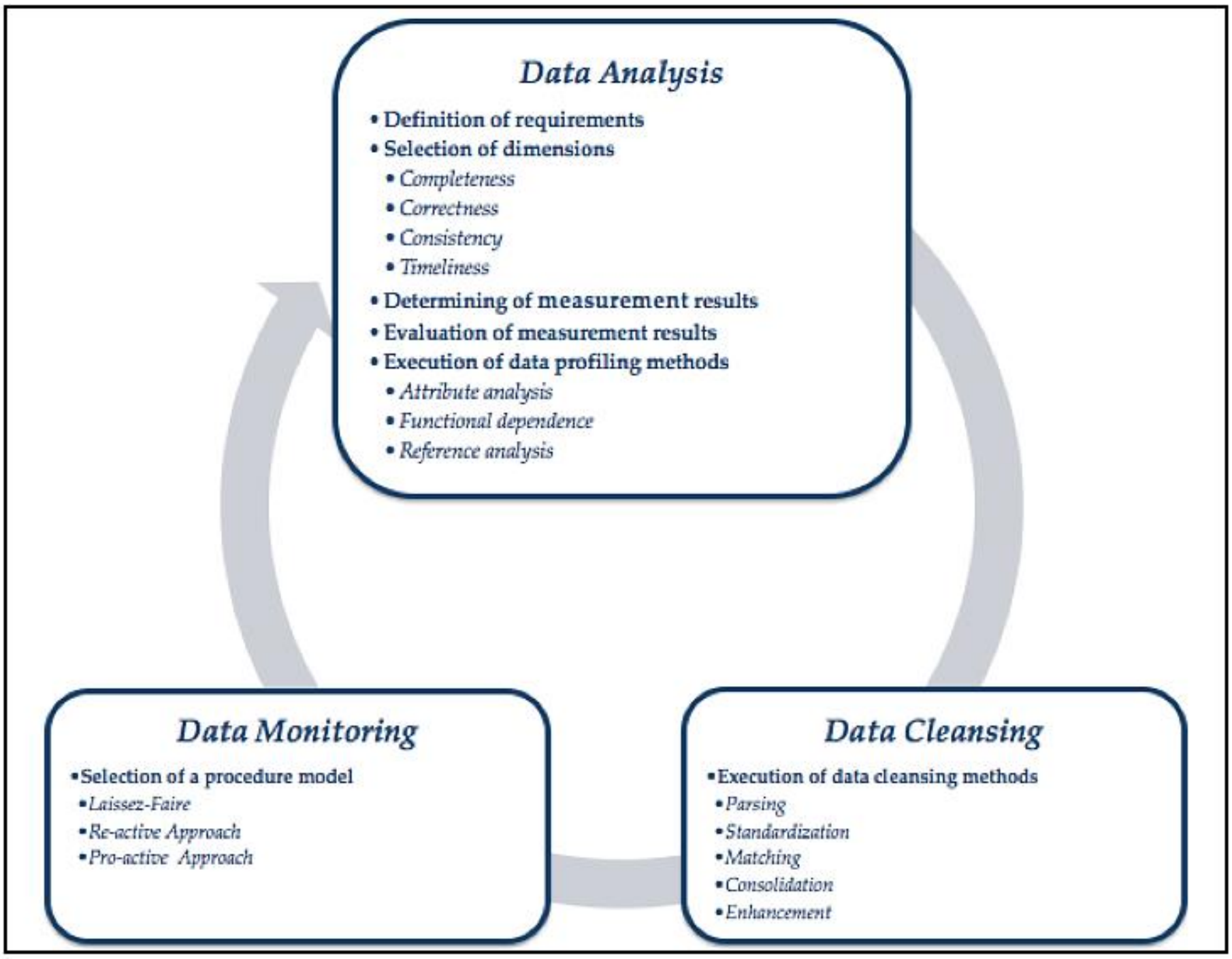

**Figure 5.** *Data quality management cycle in RIS*

The three phases capture data analysis, data cleansing and data monitoring, and together form a proven, practical approach to data governance - towards controlled data management in RIS that enables institutions to analyze, cleanse and control their research information. These phases are explained in more detail below.

In the first phase, the data quality criteria and corresponding requirements for them must be defined. Requirements can be set by different parties, e.g. especially by users of a RIS, but also by the RIS administrator. By determining the requirements, the data quality can be measured with appropriate metrics - completeness, correctness, consistency and timeliness (Azeroual, et al. 2018c), because they are easy to measure and represent a particularly representative illustration of the reporting for the users lead to an improved and transparent basis for decision-making (Azeroual, et al. 2018c). After the data quality measurements have been made in RIS, the institutions must review their measurement results. With the help of data profiling, causes of quality problems can be revealed (Azeroual, et al. 2018b). This data analysis and its methods are used to better understand the structure, relationships and content of different data sources in RIS and to detect and automatically correct potential errors (Azeroual, et al. 2018b). Data profiling is considered an important component in analyzing and improving data quality before data can be integrated into RIS (Azeroual, et al. 2018b). The institutions and libraries must use data profiling as early as possible to get an accurate picture of the state of their data. Because data profiling is the foundation for planning the most meaningful approach to correcting and harmonizing information assets.

Data cleansing can correct the identified issues (incomplete, incorrect, inconsistent, etc.) to increase the informational value and usefulness of the research information (Azeroual, et al. 2018a, d) (Azeroual and Abuosba, 2017). Using data cleansing methods and technologies such as parsing, standardization, enrichment, matching, and merging within the data cleansing process will liquidate data integration issues in RIS (Azeroual and Abuosba, 2017) (Azeroual, et al. 2018a, d). As these methods are essential for achieving and maintaining maximum data quality for more success in RIS (Azeroual, et al. 2018a, d).

If a certain data quality has been achieved, it should be preserved as long as possible. During monitoring, therefore, the data is continuously checked before being stored in the RIS because the data is constantly changing. Such long-term continuous assurance, enhancement and increase of data quality in RIS requires specific measures such as "laissez-faire", "re-active approach" and "pro-active approach" (Azeroual, et al. 2018a, d). At certain intervals, a periodic check of the entire research data should take place again and again. Even if no mistakes are made in the input of research information into RIS, relocations, mergers and other factors repeatedly result in changes to the data. Therefore, only through a permanent data control, the device is able to provide anytime information about their data quality status, as well as to increase confidence in the existing data.

To successfully ensure data quality, institutions and libraries must find a methodology and solution that includes all three key components: data analysis, data cleansing and data control. By combining this sophisticated methodology with their appropriate technology, data quality in RIS is significantly improved and increased. This will lead to sound decisions not only for information purposes and reporting but also for the creation of the information value. Than both are based on the availability of high quality data.

## 7 Conclusion

Text data mining has become progressively important as a scientific method in recent years and its potential is very high. On the one hand, most of the information is in text form and on the other, knowledge and information about research activities play a very important role in the success of an institution. The automated acquisition of research information and the associated advent of vast amounts of data make automated data analysis necessary in many areas. Likewise, the steady growth of scientific activities (such as publications, projects, etc.) and information is making computer-aided techniques of classification, clustering, NLP, information extraction, and information retrieval increasingly important. Text data mining techniques can be actively used not only by institutions, but also actively by libraries to improve their services or to develop new services. In addition, more and more tools are being developed to make the relatively complicated process of text data mining and the associated advantages usable for users without appropriate text data mining knowledge. However, a minimum level of knowledge in the area of databases is also necessary.

In summary, one can say that text analytic methods or text data mining techniques are today a component of advanced method development in institutions and libraries. The application and further development of such methods or techniques allows a long-term cost savings in the processing of service projects, addressing the increased expectations of scientific and scientific actors in the context of the establishment of open data, open science and big data, increasing the quality of research projects and the reuse already manually processed text corpora as training data of adaptive algorithms.

When research activities by institutions and libraries are mis-captured and entered into RIS, and changes and updates occur daily in the dataset, then here is the solution, a permanent and controlled process cycle (data analysis, data cleansing and data monitoring) and continuous quality management to detect and correct errors. The subject of data quality should therefore be treated as a high priority business process to guarantee and increase the value proposition of the research information produced, especially in the new environment of open science, and to reduce the efforts of the staff and to save the costs of the institutions.

## Acknowledgements

This work has been funded by the German Center for Higher Education Research and Science Studies (DZHW) and by the German Federal Ministry of Education and Research (BMBF) in the context of the project “Helpdesk to facilitate implementation of the Research Core Dataset” (http://kerndatensatz-forschung.de/).

## References

Azeroual, O. and Abuosba, M. (2017). Improving the data quality in the research information systems. *International Journal of Computer Science and Information Security*, 15(11): 82-86, November 2017.

Azeroual, O., Saake G. and Abuosba, M. (2018a). Data quality measures and data cleansing for research information systems.

*Journal of Digital Information Management*, 16(1): 12-21, February 2018.

Azeroual, O., Saake G. and Schallehn, E. (2018b). Analyzing data quality issues in research information systems via data profiling. *International Journal of Information Management,* volume 41, pages 50-56, April 2018. DOI: https://doi.org/10.1016/j.ijinfomgt.2018.02.007

Azeroual, O., Saake G. and Wastl, J. (2018c). Data measurement in research information systems: metrics for the evaluation of data quality. *Scientometrics,* volume 115, pages 1271-1290, April 2018. DOI: https://doi.org/10.1007/s11192-018-2735-5

Azeroual, O., Saake G. and Abuosba, M. (2018d). Investigations of concept development to improve data quality in research information systems (Untersuchungen zur Konzeptentwicklung für eine Verbesserung der Datenqualität in Forschungsinformationssystemen). *Proceedings of the 30th GI-Workshop on Foundations of Databases (Grundlagen von Datenbanken)*, volume 2126, pages 29-34, CEUR-WS, May 22-25, 2018, Wuppertal, Germany.

Bhanuse, S. S., Kamble, S. D., Kakde, S. M. (2016). Text mining using metadata for generation of side information. *Procedia Computer Science* 78, 807-814.

Biesenbender, S., Hornbostel, S. (2016). The research core dataset for the German science system: Developing standards for an integrated management of research information. *Scientometrics*, volume 108, pages 401–412, March 2016. DOI: https://doi.org/10.1007/s11192-016-1909-2

Brennan, N. (2018). CRIS systems as key e-infrastructure elements to support open science implementation within the European research area. In *CRIS2018: 14th International Conference on Current Research Information Systems*, Umeå, June 13-16, 2018. https://dspacecris.eurocris.org/handle/11366/695

Carstensen, K.-U., Ebert, C., Jekat, S., Klabunde, R. and Langer, H. (2010). *Computerlinguistik und Sprachtechnologie: Eine*

*Einführung*. Heidelberg: Spektrum Akademischer Verlag, Springer.

de Castro, P. (2018). The role of current research information systems (CRIS) in supporting open science implementation: the case of strathclyde. *ITLib* (2). https://dspacecris.eurocris.org/handle/11366/691

Chopra, A., Prashar, A. and Sain, C. (2013). Natural Language Processing. *International Journal of Technology Enhancements and Emerging Engineering Research*, Vol 1, Issue 4, pages 131-134, November 2013.

Clements, A. and J. Proven (2015). The emerging role of institutional CRIS in facilitating open scholarship. In *LIBER Annual Conference 2015*, London, June 25th, 2015. https://dspacecris.eurocris.org/handle/11366/393

European Commission (2017). *Evaluation of research careers fully acknowledging open science practices*. Directorate-General for Research and Innovation Open Science and ERA, Brussels. https://ec.europa.eu/research/openscience/pdf/os_rewards_wgreport_final.pdf

Heyer, G., Quasthoff, U. and Wittig, T. (2008). *Text Mining: Wissensrohstoff Text*, 1. Auflage, W3L-Verlag, Herdecke.

Liao, S.-H., Chu, P.-H. and Hsiao, P.-Y. (2012). Data mining techniques and applications - A decade review from 2000 to 2011. *Expert Systems with Application*, 39(12): 11303-11311. DOI: https://doi.org/10.1016/j.eswa.2012.02.063

MESRI (2018). *National plan for open science*. Ministère de l'Enseignement Supérieur, de la Recherche et de l'Innovation, Paris.

Schöpfel, J. (2015). The SAFARI syndrome. Implementing CRIS and open science. In *euroCRIS Membership Meeting at AMUE*, Paris, 11-12 May 2015. https://www.slideshare.net/Schopfel/safari-syndrome-48709700

Seth, S., Rüping, S., Wrobel, S. (2009). Metadata extraction using text mining. In: Solomonides, T., Hofman-Apitius, M., Freudigmann, M., Semler, S. C., Legré, Y., Kratz, M. (Eds.), *Healthgrid Research,*

*Innovation and Business Case*. IOS Press, Amsterdam, pp. 95-104.

Simons, E. (2017). Open science: the crucial importance of metadata. In *euroCRIS Strategic Membership Meeting Autumn 2017*, CVTI SR, Bratislava, Slovakia, Nov 20-22, 2017. https://dspacecris.eurocris.org/handle/11366/627

Tatum, C. (2017). What is the evaluative object of open science? In *22nd Nordic Workshop on Bibliometrics and Research Policy,* Helsinki, 09 November 2017. https://figshare.com/articles/What_is_the_evaluative_object_of_Open_Science_/5624728

Upshall, M. (2014). Text mining: Using search to provide solutions. In *Business Information Review*, 2014, Vol. 31, Nr. 2: 91–99.

## A biographical note on the authors

**M.Sc. Otmane Azeroual** is a researcher at the German Institute for Higher Education Research and Science Studies (DZHW) in Berlin, Germany. After studying Business Information Systems at the University of Applied Sciences Berlin (HTW), he began his Ph.D. in Computer Science at the Institute for Technical and Business Information Systems (ITI), Database and Software Engineering Group of the Otto-von-Guericke-University Magdeburg, Germany and at the Department of Computer Science and Engineering of the University of Applied Sciences (HTW) Berlin, Germany. His research interest is in the area of database systems, information systems, data quality management, business intelligence, big data, information security, cloud data management, open data, project management and industry 4.0.

**Prof. Dr. rer. nat. habil. Gunter Saake** is a full professor of Computer Science. He is the head of the Databases and Software Engineering Group at the Otto-von-Guericke-University Magdeburg, Germany. His research interests include database integration, tailor-made data management, database management on new hardware, and feature-oriented software product lines.

**Prof. Dr.-Ing. Mohammad Abuosba** is a full professor of Department of Computer Science and Engineering at the University of Applied Sciences (HTW) Berlin, Germany. His research areas are engineering, IT systems engineering (focus on database systems, product data management), modeling, quality management and project management.

**Dr. Joachim Schöpfel** is a senior lecturer in Library and Information Sciences at the University of Lille (France), researcher at the GERiiCO laboratory and consultant at the Ourouk consulting firm. He was Manager of the INIST (CNRS) scientific library from 1999 to 2008, head of the University of Lille department of information sciences from 2009 to 2012 and director of the French Atelier National de Reproduction des Thèses (ANRT) from 2012 to 2017. He teaches library marketing, auditing, valorization and digitization of cultural heritage collections, intellectual property and information science. His research interests are scientific information and communication, especially open access and open science, research data and grey literature. He is member of euroCRIS, of the NDLTD board of directors and of Knowledge Exchange.